# Phase Diagram of the B–BN System at Pressures up to 24 GPa: Experimental Study and Thermodynamic Analysis

Vladimir L. Solozhenko [a,*]  and  Vladimir Z. Turkevich [b]

[a] *LSPM–CNRS, Université Paris Nord, 93430 Villetaneuse, France*
[b] *Institute for Superhard Materials, National Academy of Sciences of Ukraine, Kiev, 04074 Ukraine*

**Abstract**

Phase relations in the B–BN system have been studied *ex situ* and *in situ* at pressures 2-20 GPa and temperatures up to 2800 K. The evolution of topology of the B–BN phase diagram has been investigated up to 24 GPa using models of phenomenological thermodynamics with interaction parameters derived from our experimental data on phase equilibria at high pressures and high temperatures. There are two thermodynamically stable boron subnitrides in the system i.e. $B_{13}N_2$ and $B_{50}N_2$. Above 16.5 GPa the $B_{50}N_2 \rightleftarrows L + B_{13}N_2$ peritectic reaction transforms to the solid-phase reaction of $B_{50}N_2$ decomposition into tetragonal boron (t′-$B_{52}$) and $B_{13}N_2$, while the incongruent type of $B_{13}N_2$ melting changes to the congruent type only above 23.5 GPa. The constructed phase diagram provides fundamentals for directed high-pressure synthesis of superhard phases in the B-BN system.

**Introduction**

Diamond found a wide application in modern science and technology due to its unique properties, such as extreme hardness, a high thermal conductivity, a wide band gap, a high carrier mobility, good biocompatibility, etc.[1] However, it is non-resistant to oxidation and reactive with ferrous metals. The growing demand for advanced superhard materials in cutting and shaping hard metals and ceramics have stimulated the search for novel superhard phases that are more thermally and chemically stable than diamond.

Cubic boron nitride is a superhard phase which ranks second after diamond in terms of its use in superabrasive industry.[1] Since all known binary compounds of the B-N system show higher resistance to oxygen and ferrous metals than similar carbon materials, the superhard B–N phases are expected to combine advanced mechanical properties and high thermal and chemical stability. Understanding the fundamental principles of phase formation in the B-N system requires the construction of corresponding phase *p-T* diagram which is of utmost importance in discovering and synthesizing new advanced superhard materials.

---
[*] E-mail: vladimir.solozhenko@univ-paris13.fr

Equilibrium phase diagrams of boron[2] and boron nitride[3,4,5] have been recently constructed at pressures up to 25 GPa. In contrast, phase diagram of the B–BN binary system was studied at 5 GPa only[6]; neither ambient nor higher pressures have been considered so far. According to this phase diagram, at 5 GPa the B–BN system is characterized by L + BN ⇌ $B_{13}N_2$ peritectic equilibrium at 2600 K and L ⇌ β-$B_{106}$ + $B_{13}N_2$ eutectic equilibrium at 2300 K i.e. only $B_{13}N_2$ was considered as a thermodynamically stable boron subnitride. The structure of $B_{13}N_2$ (Fig. 1a) belongs to R-3m space group ($a$ = 5.4585(8) Å, $c$ = 12.253(2) Å) and represents a new structural type produced by the distorted $B_{12}$ icosahedra linked by N–B–N chains and inter-icosahedral B–B bonds.[7,8] However, formation of another boron-rich B–N phase, $B_{50}N_2$, has been also observed in the course of crystallization from the B–BN melt at high pressure.[9] This boron subnitride belongs to a family of tetragonal boron, and its structure (space group P-4n2; $a$ = 8.8181(2) Å, $c$ = 5.0427(10) Å) has been resolved very recently.[10] It contains the $B_{12}$-icosahedra units and nitrogen atoms; each icosahedron is bonded with six nearest icosahedra, while each N atom is four-bonded and is situated in the tetrahedral cavity formed by $B_{12}$-icosahedra (Fig. 1b). Both $B_{13}N_2$ and $B_{50}N_2$ boron subnitrides are refractory,[9,11] low-compressible[8,10,12] and superhard[13,14] phases similar to other boron-rich compounds of light elements ($B_{12}C_3$, $B_{50}C_2$, $B_{12}O_2$, etc.[15]).

In the present work phase relations in the B–BN system have been studied using *ex-situ* and *in-situ* experiments at pressures up to 20 GPa and temperatures up to 2800 K. The thermodynamic analysis based on the experimental data obtained allowed us to construct equilibrium phase diagrams of the system at pressures from ambient up to 24 GPa.

**Methods**

High-pressure experiments in the 2-8 GPa range have been performed using a toroid-type apparatus with a specially designed high-temperature cell.[16] Experiments in the 8-20 GPa range have been carried out using 10 MN LPR 1000-400/50 Voggenreiter press with Walker-type module and two-stage multianvil system with 6-rams LPQ6-2400-100 Voggenreiter press. The pressure and temperature have been either directly measured (thermocouples, pressure standards, equation-of-state cross-calibration) or estimated from previously obtained calibration curves for each high-pressure apparatus. After isothermal holding time of 30–600 s at desired pressure and temperature the samples were either quenched by switching off the power or stepwise cooled down to room temperature, and slowly decompressed down to ambient pressure. In some experiments appearance of a liquid phase upon heating was detected *in situ* by electrical resistance measurements using the method described earlier.[17]

Synchrotron X-ray diffraction *in-situ* experiments at pressures 4-6 GPa and temperatures up to 2650 K have been performed using MAX80 multianvil system at F2.1 beamline, DORIS III (DESY). The experimental setup is described elsewhere.[18] Boron nitride (grade AX05, Saint-Gobain) capsules were used to isolate the reaction mixture from the graphite heater. Energy-dispersive X-ray diffraction data were collected on a Canberra solid state Ge-detector with fixed Bragg angle $\theta = 4.555(3)°$ using a white beam collimated down to $100 \times 60$ μm$^2$. The sample temperature was measured by Pt10%Rh–Pt or W5%Re–W26%Re thermocouple. The correction for the pressure effect on a thermocouple electromotive force was made using the data of Li et al.[19] Above 2200 K the power – temperature calibration curve was linearly extrapolated to the high-temperature region up to 2800 K. Pressures at different temperatures were found from the thermoelastic equation of state of highly ordered ($P_3 = 0.98 \pm 0.02$) graphite-like hexagonal boron nitride.[20]

The samples recovered from high-pressure experiments have been studied by powder X-ray diffraction (Equinox 1000 Inel diffractometers; Cu Kα and Co Kα radiation), high-resolution scanning electron microcopy (Carl Zeiss Leo Supra 50VP, Hitachi SU8010 and JEOL JSM-6700F), and energy-dispersive X-ray analysis (SDD X-Maxn EDX system, Oxford Instruments). X-ray diffraction patterns have been analyzed using Powder Cell software.[21]

Thermodynamic calculations of phase equilibria in the B–BN binary system at ambient and high pressure have been carried out from our experimental and literature data in the framework of the phenomenological thermodynamics using Thermo-Calc software.[22] The liquid phase was described using the subregular solution model,[23] and solid phases – in the framework of the Compound Energy Formalism (CEF).[24]

**Results and Discussion**

In the 2-8 GPa range quenching of B–BN melts of various compositions from 2500-2800 K usually results in formation of both $B_{13}N_2$ and $B_{50}N_2$ boron subnitrides in mixture with β-rhombohedral boron (β-$B_{106}$) and boron nitride (hBN or cBN, depending on pressure in accordance with equilibrium phase diagram of BN[5]) (Fig. 2). According to SEM-EDX data, both boron subnitrides contain only nitrogen and boron. No detectable amounts of oxygen or carbon have been observed in the recovered samples. The amount of $B_{13}N_2$ in relation to $B_{50}N_2$ increases with a higher BN content in the reaction mixture, however, synthesis of single-phase $B_{13}N_2$ seems to be impossible because its formation occurs according to the L + BN ⇆ $B_{13}N_2$ peritectic reaction. Liquid phase remains even after the completion of $B_{13}N_2$ crystallization, and upon cooling the composition of this liquid phase varies along the liquidus line to the L + $B_{13}N_2$ ⇆ $B_{50}N_2$ peritectic point and results in crystallization of $B_{50}N_2$. Further cooling leads to the L ⇆ β-$B_{106}$ + $B_{50}N_2$ eutectic reaction due to which the recovered samples always contain β-$B_{106}$.

Tetragonal boron subnitride $B_{50}N_2$ forms as a result of interaction between boron-rich liquid and rhombohedral boron subnitride $B_{13}N_2$. Thus, appearance of $B_{13}N_2$ always occur prior to formation of $B_{50}N_2$ (even for boron-rich reaction mixtures), and $B_{13}N_2$ is always present in the recovered samples, usually in mixture with boron and boron nitride.

It should be noted that chemical interaction between BN and boron may lead to the formation of a liquid by metastable eutectic reaction L ⇆ β-$B_{106}$+ BN already at temperatures of about 2400 K (at 5 GPa), and from this metastable liquid $B_{13}N_2$ can crystallize.[6]

Melting of rhombohedral boron subnitride $B_{13}N_2$ at pressures up to 8 GPa has been *in-situ* studied using synchrotron X-ray diffraction and electrical resistivity measurements.[11] The melting curve exhibits positive slope of about 30 K/GPa (see Fig. 3), which points to a lower density of the liquid phase as compared to solid $B_{13}N_2$. In all experiments crystallization of the liquid resulted in formation of $B_{13}N_2$ in mixture with boron nitride and tetragonal boron subnitride $B_{50}N_2$ that is indicative of the incongruent type of $B_{13}N_2$ melting in the 2-8 GPa pressure range.

In a special set of experiments at pressures from 15 to 20 GPa it was found that $B_{50}N_2$ undergoes solid-state decomposition of into boron (usually a mixture of $\gamma$-$B_{28}$[25] and t'-$B_{52}$[26] allotropes) and $B_{13}N_2$ already at about 2400 K i.e. at temperature distinctly below the appearance of a liquid phase in the B–BN system at these pressures.

In some rapid (by switching off the power) quenching experiments tetragonal boron subnitride $B_{50}N_2$ was synthesized in a mixture with another phase of the same composition, tetragonal $B_{48}B_2N_2$ that was previously synthesized at ambient pressure.[27] The structure of $B_{48}B_2N_2$ was described earlier by Will and Kossobutzki[28] and consists of $B_{12}$-icosahedra (each icosahedron is bonded with 10 other icosahedra), and four-bonded N atoms located in tetrahedral cavities formed by $B_{12}$-icosahedra. In contrast to $B_{13}N_2$ and $B_{50}N_2$ subnitrides, $B_{48}B_2N_2$ is characterized by high reactivity and can be easily dissolved in 7N $HNO_3$ at 370 K. This phase is considered as metastable and thus should be absent in the equilibrium phase diagram of the B–BN system.

Hypothetical $\alpha$-rhombohedral-boron-like boron subnitride $B_{38}N_6$ predicted from first principles[29] was never observed in our experiments, even below 7.5 GPa i.e. in the pressure range where it was claimed to be the only thermodynamically stable boron subnitride. All this clearly illustrate the limitations of the first-principles approach in the case of boron-rich solids, especially under pressure. Thus, only thermodynamic approach based on real phase relations from experimental data allows constructing equilibrium phase diagrams of systems containing boron.

Thermodynamic calculations of phase equilibria in the B–BN binary system have been performed based on our experimental and literature data from ambient pressure to 24 GPa with 0.5 GPa-steps. Thermodynamic data of boron allotropes, hBN and cBN were taken from.[2,3,6] Molar volumes ($V_0$), bulk moduli ($B$), their first pressure derivatives ($B'$), and volume thermal expansion coefficients ($\alpha_v$) for liquid phase, $\beta$-$B_{106}$, $\gamma$-$B_{28}$ and t'-$B_{52}$ were taken from[2,31], for hBN and cBN – from[18,32,33], and for $B_{50}N_2$ and $B_{13}N_2$ – from[10,12,34]. Pressure dependencies of molar volumes were represented using the Murnaghan approximation.[30]

The molar volume of the liquid phase was described by the equation:

$$V_L = V_B x_B + V_{BN} x_{BN} + \Delta V^{mix} x_B x_{BN},$$

where $\Delta V^{mix} = 0.5 \times 10^{-6}$ cm$^3$/mole is mixing volume. The value of mixing volume and expression for Gibbs free energy of $B_{13}N_2$ i.e. $G_{B13N2} = 13 G_B + 2 G_{BN} - 170000 + 60 \times T$ J/mole were determined by the solution of inverse problem using experimental data on $B_{13}N_2$ melting. Pressure dependence of temperature of the L + BN ⇌ $B_{13}N_2$ peritectic reaction calculated in the framework of our thermodynamic approach is presented in Fig. 3 by dashed line. One can see that this line is in a good agreement with two sets of experimental data on $B_{13}N_2$ melting.

Since data on $B_{50}N_2$ formation enthalpy are not available, in the present study the expression $G_{B50N2} = 48G_B + 2G_{BN} - 315000 + 100 \times T$ J/mole was used that allowed us to fix at 2770 K the temperature of the $t'\text{-}B_{52} + B_{13}N_2 \rightleftarrows B_{50}N_2$ peritectoid equilibrium observed in our experiments at 17 GPa.

At lower pressures, say at 6 GPa, the existence of two peritectic equilibria: $L + BN \rightleftarrows B_{13}N_2$ at 2610 K and $L + B_{13}N_2 \rightleftarrows B_{50}N_2$ at 2520 K allows us to explain the fact that upon cooling crystallization from the B–BN melt of $B_{50}N_2$ composition results first in formation of $B_{13}N_2$, which further (due to the interaction with boron-rich liquid) forms $B_{50}N_2$. Finally, cooling results in formation of $B_{50}N_2$ and $\beta\text{-}B_{106}$ mixture according to the $L \rightleftarrows \beta\text{-}B_{106} + B_{50}N_2$ eutectic reaction at 2440 K.

The pressure evolution of the B–BN phase diagram is shown in Fig. 4. In addition to the quantitative changes of the diagram parameters (equilibria temperatures and limiting solubilities), variations of the diagram topology are observed, i.e. above 8.5 GPa the $\gamma\text{-}B_{28} \rightleftarrows t'\text{-}B_{52}$ equilibrium line appears; and the incongruent type of $B_{13}N_2$ melting transforms to the congruent one at about 23.5 GPa, i.e. the $L + BN \rightleftarrows B_{13}N_2$ peritectic changes to the $L \rightleftarrows B_{13}N_2 + BN$ eutectic. At pressures below 16.5 GPa $B_{50}N_2$ melts according to the $L + B_{13}N_2 \rightleftarrows B_{50}N_2$ peritectic reaction. Above 16.5 GPa the mentioned peritectic reaction transforms to the solid-phase decomposition of $B_{50}N_2$ into $t'\text{-}B_{52}$ and $B_{13}N_2$. Further pressure increase leads to the decreasing temperature of the $B_{50}N_2 \rightleftarrows t'\text{-}B_{52} + B_{13}N_2$ peritectoid reaction from 2740 K at 16.5 GPa to 2330 K at 24 GPa.

**Conclusions**

Chemical interaction and phase relations in the B–BN system have been studied in *ex-situ* and *in-situ* experiments at pressures 2-20 GPa and temperatures up to 2800 K. Based on the obtained experimental data we have constructed equilibrium phase diagram of the B–BN system and studied its pressure evolution up to 24 GPa. It was found that there are two thermodynamically stable boron subnitrides in the system, i.e. rhombohedral $B_{13}N_2$ and tetragonal $B_{50}N_2$. Above 16.5 GPa the $B_{50}N_2 \rightleftarrows L + B_{13}N_2$ peritectic reaction transforms to the $B_{50}N_2 \rightleftarrows t'\text{-}B_{52} + B_{13}N_2$ peritectoid reaction, while the incongruent type of $B_{13}N_2$ melting changes to the congruent type only above 23.5 GPa. The constructed equilibrium phase diagram opens a way for directed synthesis of all compounds of the B–BN system at pressures up to 24 GPa and for high-pressure design of new advanced B–N materials.

**Acknowledgments**

The authors thank Drs. Vladimir A. Mukhanov and Kirill A. Cherednichenko for assistance in high-pressure experiments and Dr. Volodymyr Bushlya for SEM-EDX analysis. Synchrotron X-ray diffraction experiments at DESY have been performed during beam time allocated to Project DESY-D-I-20090007 EC, and received funding from the European Community's Seventh Framework Programme (FP7/2007-2013) under grant agreement No 226716. The 6-rams press experiments were carried out at P61.2 beamline, PETRA III Extension during time kindly provided by DESY; assistance of Dr. Norimasa Nishiyama is greatly acknowledged. This work was financially supported by the European Union's Horizon 2020 Research and Innovation Programme under the Flintstone2020 project (grant agreement No 689279).

**ORCID IDs**

Vladimir L. Solozhenko  https://orcid.org/0000-0002-0881-9761

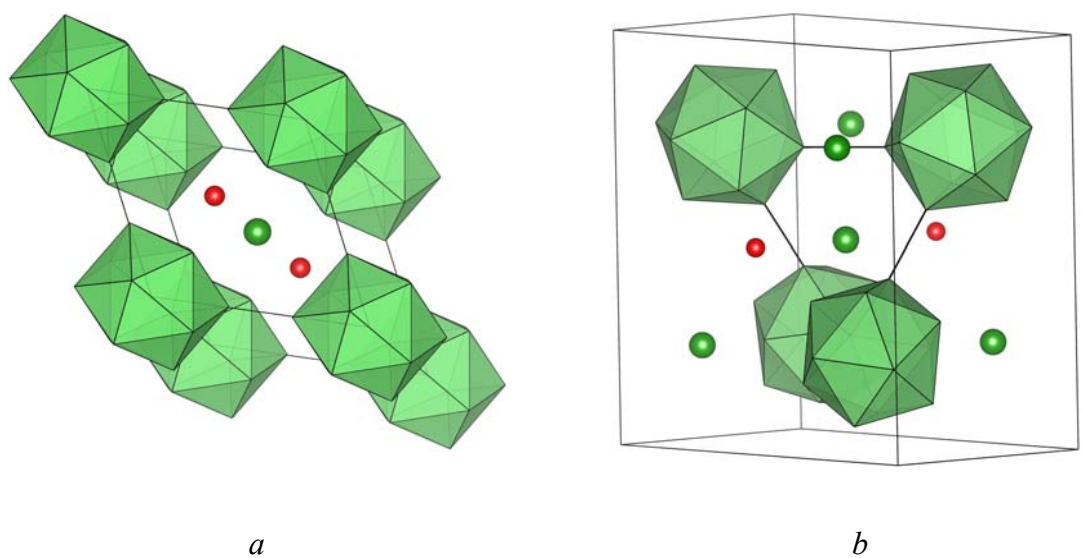

Fig. 1 Crystal structures of rhombohedral $B_{13}N_2$ [7] (*a*) and tetragonal $B_{50}N_2$ [10] (*b*). Nitrogen and boron atoms are presented in red and green, respectively.

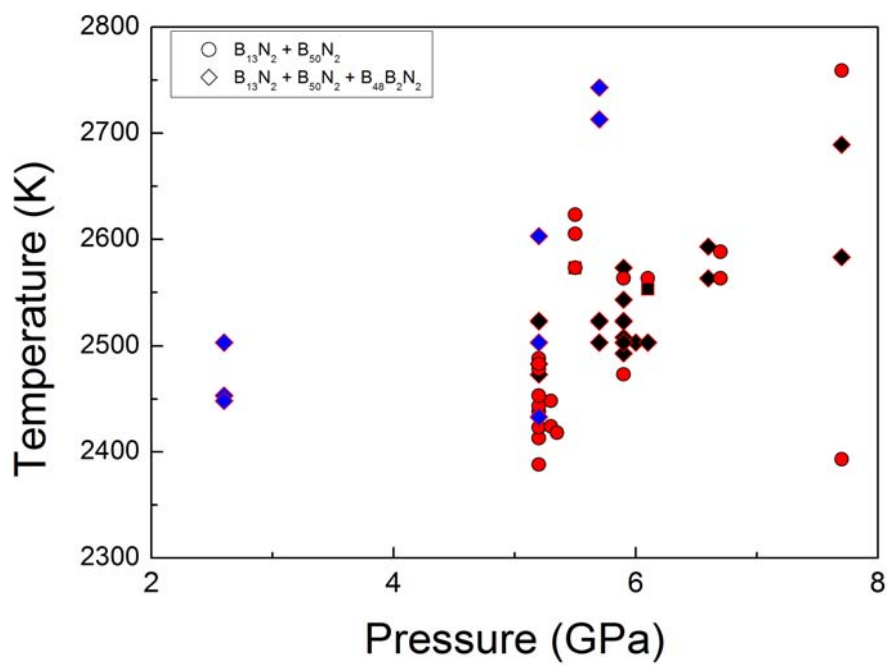

Fig. 2  Results of *ex-situ* studies of phase formation in the B–BN system by crystallization from melt for reaction mixtures of $B_{13}N_2$ (red and black symbols) and $B_{50}N_2$ (blue symbols) compositions. All samples also contain detectable amounts of β-$B_{106}$ and boron nitride (hBN and/or cBN).

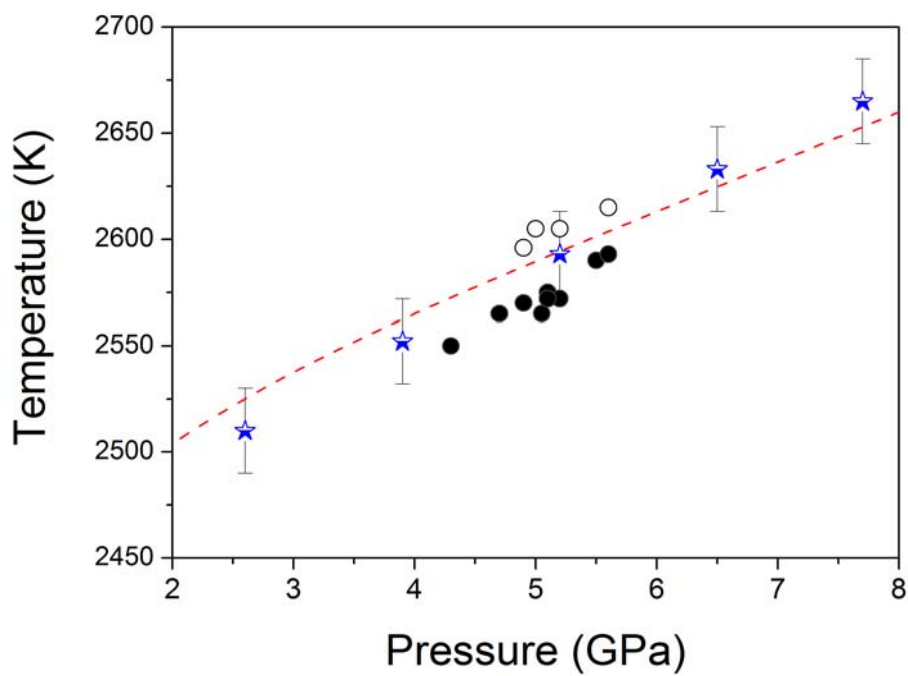

Fig. 3  $B_{13}N_2$ melting temperature *versus* pressure. The results of synchrotron X-ray diffraction experiments are presented by circles (solid symbols correspond to solid $B_{13}N_2$, open symbols – to the liquid). Half-filled stars show the melting onset registered *in situ* by electrical resistivity measurements;[11] dashed line is calculated pressure dependence of temperature of the L + BN ⇌ $B_{13}N_2$ peritectic reaction.

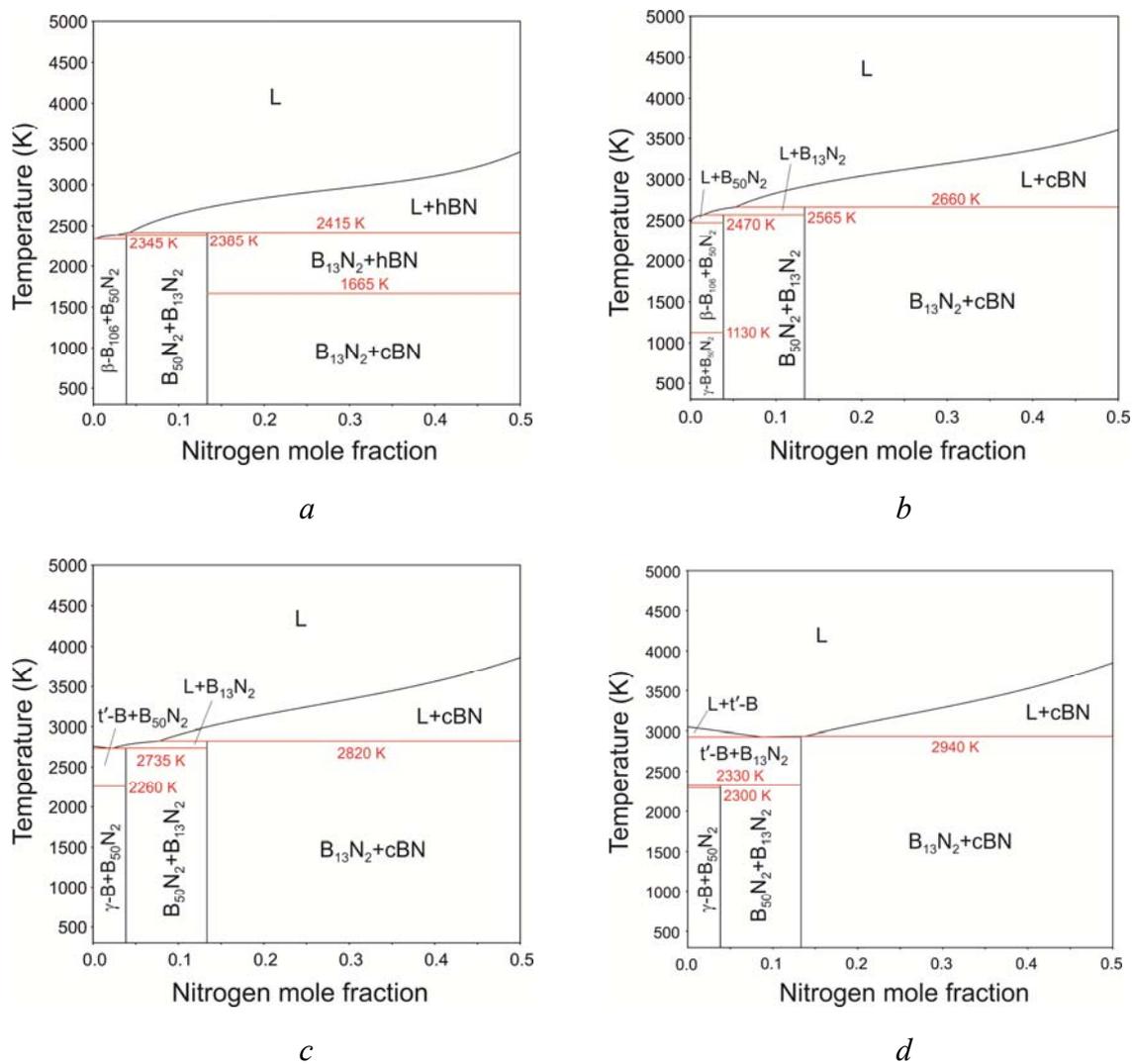

Fig. 4  Phase diagram of the B–BN system at 0.1 MPa (*a*), 8 GPa (*b*), 16 GPa (*c*) and 24 GPa (*d*); β, γ, t′ are boron allotropes (β-$B_{106}$, γ-$B_{28}$ and t′-$B_{52}$, respectively).